\documentclass[journal]{IEEEtran}
\usepackage{amsfonts}
\usepackage{amsmath}
\usepackage{amsthm}
\usepackage{amssymb}
\usepackage{graphicx}
\usepackage[T1]{fontenc}
\usepackage{supertabular}
\usepackage{longtable}
\usepackage[usenames,dvipsnames]{color}
\usepackage{bbm}
\usepackage{fancyhdr}
\usepackage{breqn}
\usepackage{fixltx2e}
\usepackage{capt-of}
\usepackage{mdframed}
\newcommand{\mathsym}[1]{{}}
\newcommand{\unicode}[1]{{}}

\begin{document}

\title{\color{Brown}How to (Not) Estimate Gini Coefficients for Fat Tailed Variables}
\author{ 
\IEEEauthorblockN{Nassim Nicholas Taleb
} \\
    } 
\maketitle

\begin{abstract}
Direct measurements of Gini coefficients by conventional arithmetic calculations are a poor estimator, even if paradoxically, they include the entire population, as because of super-additivity they cannot lend themselves to comparisons between units of different size, and intertemporal analyses are vitiated by the population changes. The Gini of aggregated units A and B will be higher than those of A and B computed separately. This effect becomes more acute with fatness of tails. When the sample size is smaller than entire population, the error is extremely high. We compare the standard methodologies to the indirect methods via maximum likelihood estimation of tail exponent. The conventional literature on Gini coefficients cannot be trusted and comparing countries of different sizes makes no sense; nor does it make sense to make claims of "changes in inequality" based on such measure.\\
\indent We compare to the tail method which is unbiased,  with considerably lower error rate.  We also consider measurement errors of the tail exponent and suggest a simple but efficient methodology to calculate Gini coefficients.

\end{abstract}

\thanks{}
\thispagestyle{fancy} 

\thanks{}
\markboth{\textbf{Tail Risk Working Paper Series}}
\flushbottom

\begin{mdframed}
	
\section{Introduction/Summary}
Consider 10 separate countries, cities, or other units of equal size, with population $10^3$. Assume the wealth in each unit follows a power law distribution, say a Pareto-Lomax, all with the exact same parameters. Assume a tail exponent of 1.1. The average Gini coefficient as obtained by direct measurement will be $\approx .71$ per country. Now aggregate them into a single country. The composite Gini ---as traditionally and currently measured --- will be $\approx .75$, that is 6\% higher --for the \textit{same} sample. This inconsistency implies not only that the Gini cannot lend itself to comparisons between units of different size but that intertemporal assessments are vitiated by the population changes.

Further, the sampling error remains high throughout.

The effect is similar to the one about percentile in \cite{taleb2015super}.

This note shows that Gini Coefficient by direct measurement as estimator is not consistent, downward biased and lends itself to illusions; maximum likelihood (ML) parametrization of tail exponent is more efficient, unbiased, and economical of data: its error rate can be more than one order of magnitude smaller than the "direct" gini measurement. We get explicit distributions for the maximum likelihood estimator.
\end{mdframed}

Table \ref{tableofresults} presents our story and its conclusion; it compares the Gini coefficient obtained by conventional arithmetic calculations to the Maximum Likelihood estimation via tail exponent.  We ran Monte Carlo simulations ($10^8$) for a Pareto distribution with exponent $\alpha=1$. For the first category, "direct", we estimated the Gini using conventional methods. For the second Maximum Likelihood (ML) we estimated the tail exponent and expressed the resulting Gini.

shows how unbiased and error prone the Gini coefficient is from direct measurement for different population size.

\begin{table}[h]
\begin{center}
\caption{Comparison of direct Gini to ML estimator, assuming tail $\alpha=1.1$}\label{tableofresults}
\label{biases}%
\begin{tabular}{c|ccc|cc|c}
$n$ & &\textbf{Direct} & & & \textbf{ML} \\
(popul & Mean & Bias & STD & Mean & STD & Error \\ 
or sample)&  &  &  & & & ratio\\ \hline
$10^3$ & 0.711 & -0.122&  0.0648  &.8333 & 0.0476 & 1.4\\ 
$10^4$ &0.750 & -0.083 & 0.0435   &.8333  &0.015 & 3\\ 
$10^5$ & 0.775 & -0.058& 0.0318   & .8333 & 0.0048 & 6.6\\ 
$10^6$ & 0.790&-0.043 &  0.0235 &.8333 & 0.0015 & 156\\ 
$10^7$ & 0.802&  -0.033& 0.0196  &.8333 & $\approx 0$& $ > 10^5$\\
\end{tabular}%
	
\end{center}

\end{table}

\section{Statement}
\begin{figure}[h!]
\includegraphics[width=\linewidth]{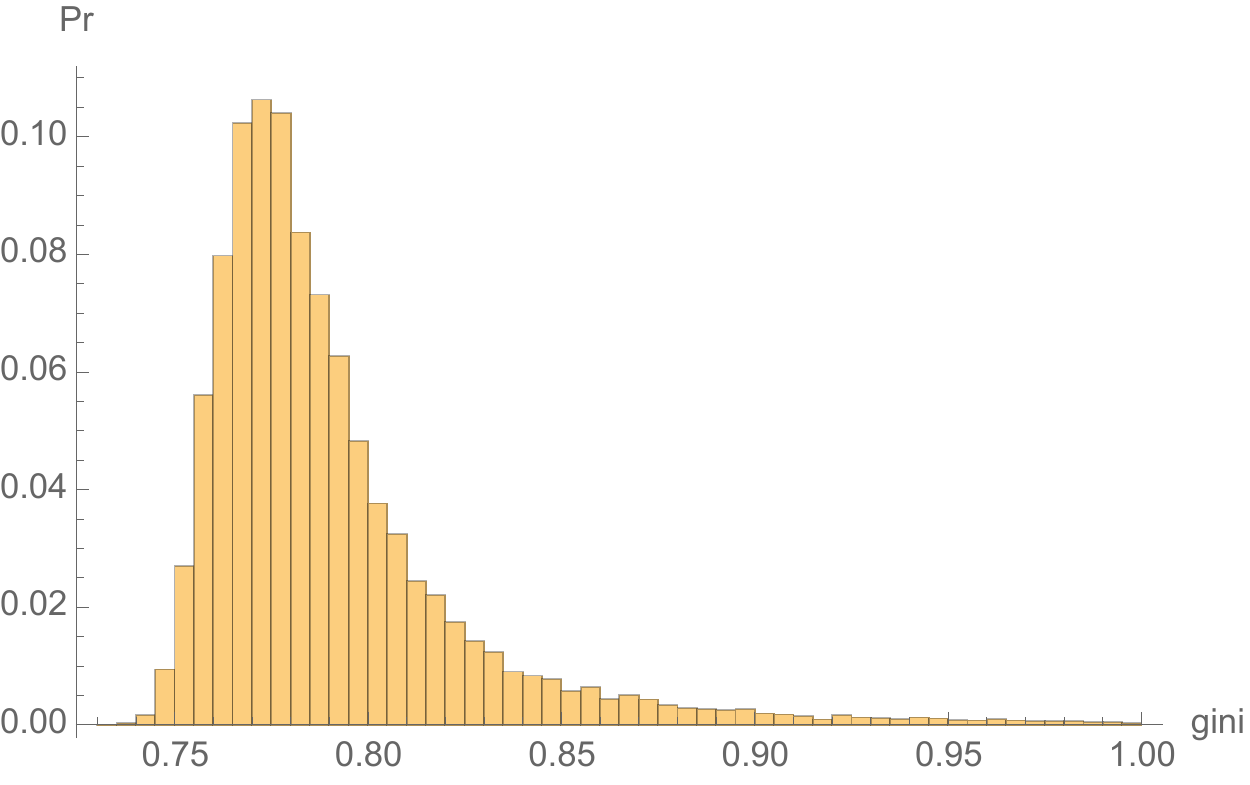}
\caption{Histogram of the distribution of direct estimation, population $=10^6$. We notice a long right tail bounded at $1$.}
\includegraphics[width=\linewidth]{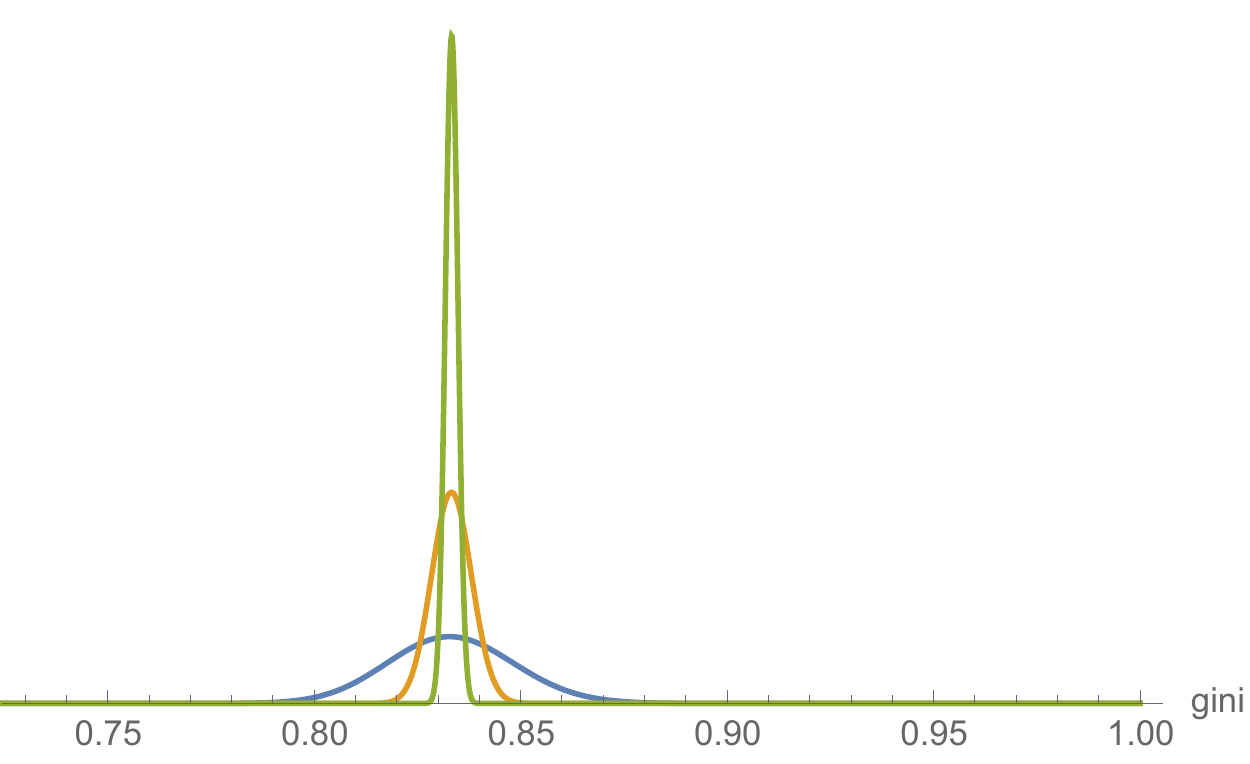}
\caption{Distribution of indirect estimator via exponent $n=10^4, 10^5,10^6$.}
\end{figure}

Where $g$ is the Gini coefficient and $X$ and $X'$ are independent (etc., etc.) with mean $\mu$:
\begin{equation}
g=\frac{1}{2}\frac{\mathbb{E}\left(|X-X'|\right)}{\mu}. \label{GINI}
\end{equation}
In other words the Gini is the mean expected deviation between any two random variables ("mean difference") scaled by the mean.

We can express the Gini as half the relative mean difference, where sample $Y=(Y_i)_{1\leq i \leq n}$, with apparent estimator:
\begin{equation}
	\widehat{G}_d(Y)=\frac{\sum _{j=1}^n \sum _{i=1}^n \left| Y_i-Y_j\right| }{2 (n-1) \sum _{i=1}^n Y_i}
\end{equation}
which can be further simplified 
\begin{equation}
\widehat{G}_d(Y)=-\frac{1}{n}\left(\frac{2 \sum _{i=1}^n (n-i+1) Y_{(i)}}{\sum _{i=1}^n Y_{(i)}}+n+1	\right)
\end{equation}
where $X_{(1)},X_{(2)},..., X_{(n)}$ are the ordered statistics of $X_1,...,X_n$ such that: $X_{(1)} < X_{(2)} < ... < X_{(n)}$.

For a power law distribution, $\widehat{G(Y)}$ is a slowly converging estimator, downward biased, inconsistent under aggregation, so, with $n_Y$ and $n_X$ the relative sample sizes of $X$ and $Y$ respectively. 
We conjecture that, for $X$ and $Y$ following the same distribution:
	\begin{equation}
	 \widehat{G}(Y \sqcup X) \geq \frac{n_Y}{n_X+n_Y} \widehat{G}(Y) + \frac{n_X}{n_X+n_Y} \widehat{G}(X).
\end{equation}

 This inequality is derived in a similar way  to the inequality in theorem 1 in \cite{taleb2015super}.

Next we show that for power law an "indirect" estimation via the Hill estimator of the tail exponent is a more efficient way to estimate the Gini coefficient.

\section{Estimation from tail $\alpha$ via Maximum Likelihood}
This section finds explicit distributions for the estimator.
\subsection{Power Law Gini Coefficient}
If we know the distribution of X, then Equation \ref{GINI} is straightforward.
In the event of known cumulative distribution function $\Phi$, consider that $|X-X'|=X+X'-2 \,\text{min}(X, X')$. Hence the expectation becomes:
$$\mathbb{E}\left(|X-X'|\right)=2 \left(\mu - \mathbb{E}(X,X')^-\right)$$
We have the joint cumulative
$$F\left((x,x')^-\right)= 1-\mathbb{P}(X>x) \mathbb{P}(X'>x)$$
hence, with $X \in [L,\infty)$:
\begin{equation}
G=1-\frac{1}{\mu}\int_L^\infty \left(1- \Phi(x)\right)^2 \, \mathrm{d} x \label{ginsquares}
\end{equation}


\subsection{Distribution of the exponent}

Next we calculate the distribution of the tail exponent of a power law. We start with the standard Pareto distribution for random variable $X$ with pdf:

\begin{equation}
\phi_X(x)=\alpha  L^{\alpha } x^{-\alpha -1}	\,, x>L
\end{equation}

Assume $L=1$ by scaling.

The likelihood function is
$\mathcal{L}=\prod _{i=1}^n \alpha  x_i^{-\alpha -1}$. Maximizing the Log of the likelihood function (assuming we set the minimum value) $ \log(\mathcal{L})=n (\log (\alpha )+\alpha  \log (L))-(\alpha +1) \sum _{i=1}^n \log \left(x_i\right)$ yields: $\hat{\alpha} = \frac{n}{\sum _{i=1}^n \log \left(x_i\right)}.$ Now consider $l=-\frac{\sum _{i=1}^n \log  X_i}{n}$. Using the characteristic function to get the distribution of the average logarithm yield:
$$\psi(t)^n=\left(\int_1^{\infty } f(x) \exp \left(\frac{i t \log (x)}{n}\right) \, dx\right)^n=\left(\frac{\alpha  n}{\alpha  n-i t}\right)^n$$ which is the characteristic function of the gamma distribution $(n,\frac{1}{\alpha  n})$. A standard result is that $\hat{\alpha}' \triangleq \frac{1}{l}$ will follow the inverse gamma distribution with density: $$\phi_{\hat{\alpha}}(a)=\frac{e^{-\frac{\alpha  n}{\hat{\alpha }}} \left(\frac{\alpha  n}{\hat{\alpha }}\right)^n}{\hat{\alpha } \Gamma (n)}, \, a>0$$.
\subsubsection{Debiasing} 
Since $\mathbb{E}(\hat{\alpha})=\frac{n}{n-1} \alpha$ we elect another --unbiased-- random variable $\hat{\alpha'}= \frac{n-1}{n} \hat{\alpha}$ which, after scaling, will have for distribution $\phi_{\hat{\alpha'}}(a)=\frac{e^{\frac{\alpha -\alpha  n}{a}} \left(\frac{\alpha  (n-1)}{a}\right)^{n+1}}{\alpha  \Gamma (n+1)}$.
\subsubsection{Truncating for $\alpha>1$}
Given that values of $\alpha \leq 1$ lead to infinite mean (hence no Gini) we restrict the distribution to values greater than $1+\epsilon$. 
Our sampling now applies to lower-truncated values of the estimator, those strictly greater than 1, with a cut point $\epsilon >0$, that is, 
 ${\sum \frac{n-1}{\log(x_i)}>1+\epsilon}$, 
or $\mathbb{E}(\hat{\alpha}|_{\hat{\alpha}>1+\epsilon})$:
$\phi_{\hat{\alpha''}}(a)=\frac{\phi_{\hat{\alpha'}}(a)}{\int_{1+\epsilon}^\infty \phi_{\hat{\alpha'}}(a) \,\mathrm{d}a}$, hence the distribution of the values of the exponent conditional of it being greater than $1$ becomes:
\begin{equation}
\phi_{\hat{\alpha''}}(a)= \frac{e^{\frac{\alpha  n^2}{a-a n}} \left(\frac{\alpha  n^2}{a (n-1)}\right)^n}{a \left(\Gamma (n)-\Gamma \left(n,\frac{n^2 \alpha }{(n-1) (\epsilon +1)}\right)\right)}	\, ,\, a \geq 1+\epsilon 
\end{equation}
\subsection{The distribution of the $\alpha$-derived Gini}

Now define the "derived gini" from estimated $\alpha$, $G \triangleq \frac{1}{2\hat{\alpha''}-1}$. After some manipulation, we have $\phi_G(g)$ the distribution of the derived gini:

\begin{multline}
\phi_G(g)= \frac{2^n e^{-\frac{2 \alpha  g n^2}{(g+1) (n-1)}} \left(\frac{\alpha  g n^2}{(g+1) (n-1)}\right)^n}{g (g+1) \left(\Gamma (n)-\Gamma \left(n,\frac{n^2 \alpha }{(n-1) (\epsilon +1)}\right)\right)} ,\\  g\in (0,\frac{1}{2 \epsilon +1})
\end{multline}

\begin{figure}
\includegraphics[width=\linewidth]{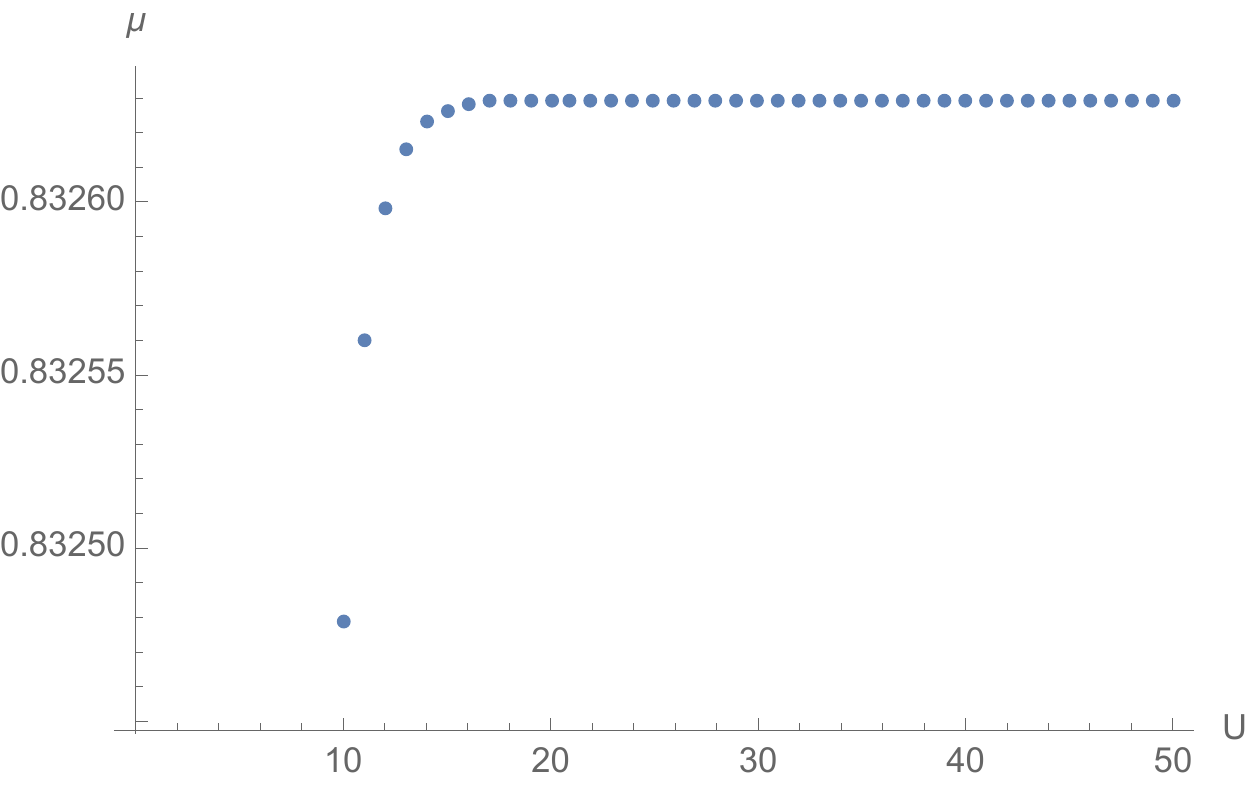}
\caption{convergence of the coefficient with number of summands $U$}\label{U}
\end{figure}

\begin{figure}
\includegraphics[width=\linewidth]{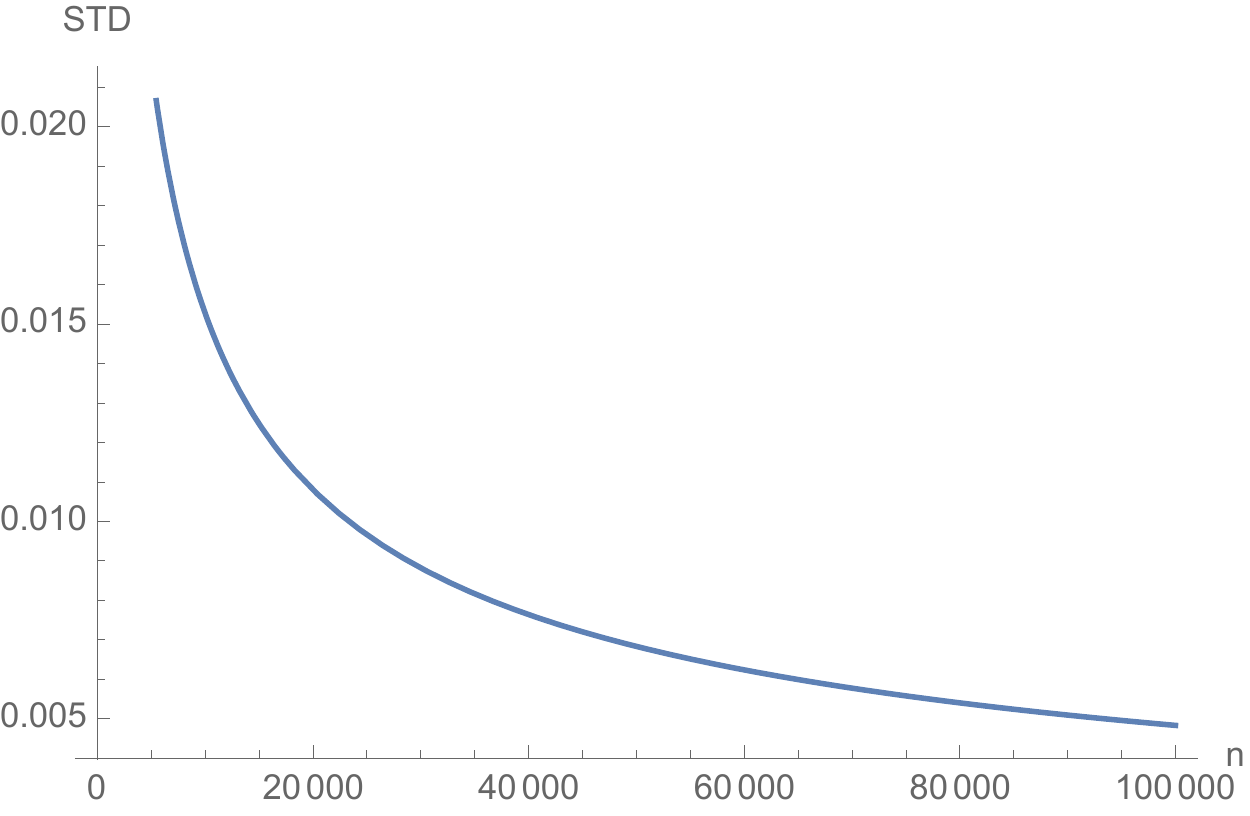}
\caption{Standard deviation of the ML estimator with an increase of population}\label{declineSTD}
\end{figure}

\subsection{Moments of the estimated Gini coefficient}

We are looking for moment of order $m$, that is $\mu(m)$ as $\int_0^ {\frac{1}{2 \epsilon +1}} g^m \phi_G(g) dg $. 
By substitution, with $u= \frac{g}{g+1}$,
$$\mu(m)=\int_0^{\frac{1}{2 \epsilon +2}} \frac{2^n \left(\frac{1}{1-u}\right)^m u^{m-1} e^{-\frac{2 \alpha  n^2 u}{n-1}} \left(\frac{\alpha  n^2 u}{n-1}\right)^n}{\Gamma (n)-\Gamma \left(n,\frac{n^2 \alpha }{(n-1) (\epsilon +1)}\right)} \, du$$
using the property that $\sum _{i=0}^{\infty } u^i \binom{i+m-1}{i}=(1-u)^{-m}$ and that 
\small
\begin{multline}
\int_0^{\frac{1}{2 \epsilon +2}} \frac{2^n u^i u^{m-1} e^{-\frac{2 \alpha  n^2 u}{n-1}} \left(\frac{\alpha  n^2 u}{n-1}\right)^n}{\Gamma (n)-\Gamma \left(n,\frac{n^2 \alpha }{(n-1) (\epsilon +1)}\right)} \, du= \left(\frac{1}{2 \epsilon +2}\right)^{i+m} \\
\frac{\left(\frac{\alpha  n^2}{(n-1) (\epsilon +1)}\right)^{-i-m} \left(\Gamma (i+m+n)-\Gamma \left(i+m+n,\frac{n^2 \alpha }{(n-1) (\epsilon +1)}\right)\right)}{\Gamma (n)-\Gamma \left(n,\frac{n^2 \alpha }{(n-1) (\epsilon +1)}\right)}	
\end{multline}
\normalsize
we finally have, with $U$ a natural number:
\begin{multline}
	\mu(m)=
	\lim_{U \to +\infty}\sum _{i=0}^U \frac{\binom{i+m-1}{i} \left(\left(\frac{1}{2 \epsilon +2}\right)^{i+m} \left(\frac{\alpha  n^2}{(n-1) (\epsilon +1)}\right)^{-i-m}\right)}{\Gamma (n)-\Gamma \left(n,\frac{n^2 \alpha }{(n-1) (\epsilon +1)}\right)}\\
	 \left(\Gamma (i+m+n)-\Gamma \left(i+m+n,\frac{n^2 \alpha }{(n-1) (\epsilon +1)}\right)\right)
\end{multline}
\normalsize
which, in practice, with values of $U \approx 7$ produces appropriate approximations, see Figure \ref{U}. We get explicit (rather, semi-explicit) expressions of the standard deviations and show their decline in Figure \ref{declineSTD}.


\subsection{Some comments}

%
%
%
%

For recent wealth data restating Pareto and Mandelbrot's point \cite{mandelbrot1960pareto}, see \cite{dagsvik2013pareto}. Some authors missed the point: see   \cite{lerman1989improving}, \cite{gastwirth1976interpolation}, \cite{alvaredo2011note}.  In some cases, some get it backwards, getting $\alpha$ from $G$ \cite{wildman2003health}.
\section*{Acknowledgment}
Raphael Douady, Pasquale Cirillo.

\bibliographystyle{IEEEtran}
\bibliography{/Users/nntaleb/Dropbox/Central-bibliography}

\end{document}